\documentclass[conference]{IEEEtran}

\usepackage{graphicx}
\usepackage{subfigure}
\usepackage{amsmath}
\usepackage{amsfonts}
\usepackage{amssymb}
\usepackage{latexsym}
\usepackage{moreverb}
\usepackage{booktabs}
\usepackage{times}
\usepackage{indentfirst}
\usepackage{algorithmic}
\usepackage{algorithm}
\usepackage[table]{xcolor}  
\usepackage{multirow}       
\usepackage{array}
\usepackage{picins}
\usepackage{wrapfig}
\usepackage{fancyhdr}

\graphicspath{{eps/}}

\newcommand{\Fig}[1]{Fig.~\ref{#1}}

\newcommand{\Figss}[1]{Figs.~\ref{#1}}

\newcommand{\Sec}[1]{Section~\ref{#1}}

\newtheorem{dfn}{Definition}

\DeclareMathOperator*{\argmin}{arg\,min}

\fancypagestyle{IEEE_Green_open_access_footer}{%
  \fancyhf{}
  
  \fancyfoot[C]{\footnotesize \copyright2016 IEEE. Personal use of this material is permitted. Permission from IEEE must be obtained for all other uses, in any current or future media, including reprinting/republishing this material for advertising or promotional purposes, creating new collective works, for resale or redistribution to servers or lists, or reuse of any copyrighted component of this work in other works. DOI: 10.1109/GLOCOM.2016.7842319}
}

\begin{document}


\title{The Privacy Exposure Problem in Mobile Location-based Services}

\author{Fang-Jing~Wu$^{1}$, Matthias~R.~Brust$^3$, Yan-Ann~Chen$^{4}$, and Tie~Luo$^2$\\
$^1$NEC Laboratories Europe, Heidelberg, Germany\\
$^2$Institute for Infocomm Research, A*STAR, Singapore\\
$^3$Singapore University of Technology and Design, Singapore\\
$^4$Department of Computer Science, National Chiao Tung University, Taiwan\\
}

\maketitle

\thispagestyle{IEEE_Green_open_access_footer}

\begin{abstract}
Mobile location-based services (LBSs) empowered by mobile
crowdsourcing provide users with context-aware intelligent
services based on user locations. As smartphones are capable of
collecting and disseminating massive user location-embedded
sensing information, privacy preservation for mobile users has
become a crucial issue. This paper proposes a metric called
\emph{privacy exposure} to quantify the notion of privacy, which
is subjective and qualitative in nature, in order to support
mobile LBSs to evaluate the effectiveness of privacy-preserving
solutions. This metric incorporates \emph{activity coverage} and
\emph{activity uniformity} to address two primary privacy threats,
namely {\em activity hotspot} disclosure and {\em activity
transition} disclosure. In addition, we propose an algorithm to
minimize privacy exposure for mobile LBSs. We evaluate the
proposed metric and the privacy-preserving sensing algorithm via
extensive simulations. Moreover, we have also implemented the
algorithm in an Android-based mobile system and conducted
real-world experiments. Both our simulations and experimental
results demonstrate that (1) the proposed metric can properly
quantify the privacy exposure level of human activities in the
spatial domain and (2) the proposed algorithm can effectively
cloak users' activity hotspots and transitions at both high and
low user-mobility levels.
\end{abstract}


\section{Introduction}

Mobile location-based services (LBSs) \cite{Rao2003_MobileLBS}
exploit location information of mobile users to provide
context-aware and personalized services. Mobile crowdsourcing,
enabled by sensor-rich and widely-used smartphones, has further
spurred a wealth of mobile LBSs such as transportation services
\cite{lau11ccsa}, lifestyle enhancement systems \cite{Yelp}, and
localization applications \cite{Wu2015_ICC}. Such mobile LBSs
highly rely on \emph{location-embedded information} from diverse
sources such as built-in sensors on mobile devices, human inputs,
and social media
\cite{chen2013modeling}\cite{chen2014game}\cite{chen2014modeling}.
This presents significant privacy threats
\cite{Barkuus2003_PrivacyforLBS} to mobile users. When they query
or contribute data to mobile LBS systems, the privacy-sensitive
information of users might be disclosed to undesired parties by
untrustworthy or poorly-designed back-end data management systems.

\begin{figure}
\centering
\includegraphics[width=0.85\columnwidth]{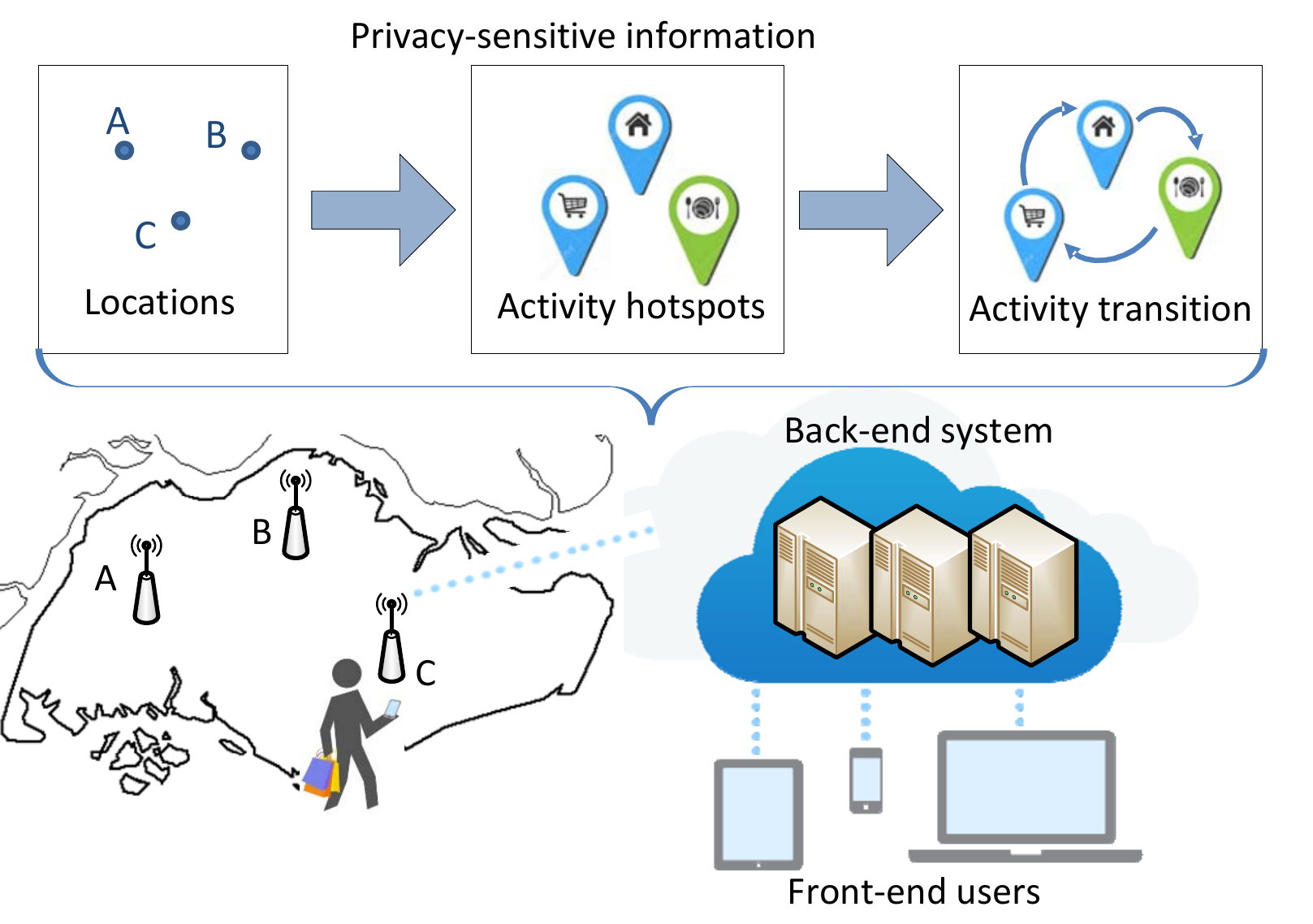}
\caption{Privacy-sensitive information in a location-based WiFi
advisory system.} \label{Fig:system-architecture}
\end{figure}

Specifically, the privacy-sensitive information in this paper
includes ``activity hotspots'' and ``activity transitions''. The
former refers to places that a user stays for a long time (e.g.,
home and office) or frequently visits (e.g., a cafe), and the
latter refers to the sequence of those activity hotspots visited
by a user. \Fig{Fig:system-architecture} illustrates a
location-based WiFi advisory system \cite{wifiscout14mass}, where
a user may query for nearby WiFi access points with good signal
quality, or contributes her quality of experience (QoE) of WiFi
usage to the system. In either case, the user may unintentionally
disclose her home and office locations as well as the travel path
between them. Such a problem becomes more thorny due to the recent
fast advancement of machine learning and data mining techniques,
where activity hotspots and transitions could be revealed through
statistics and probabilistic models
\cite{Baratchi2014_mobilityModel}\cite{Rossi2014_IdentifyingUsers},
and the ``next place'' of a user might be inferred
\cite{Scellato2011_NextPlace}\cite{Gambs2012_mobilityModel}.
Therefore, privacy preservation for mobile location-based services
and crowdsourcing applications is critical for such system to be
widely
and practically adopted. 

\begin{figure}
\centering
\includegraphics[width=\columnwidth]{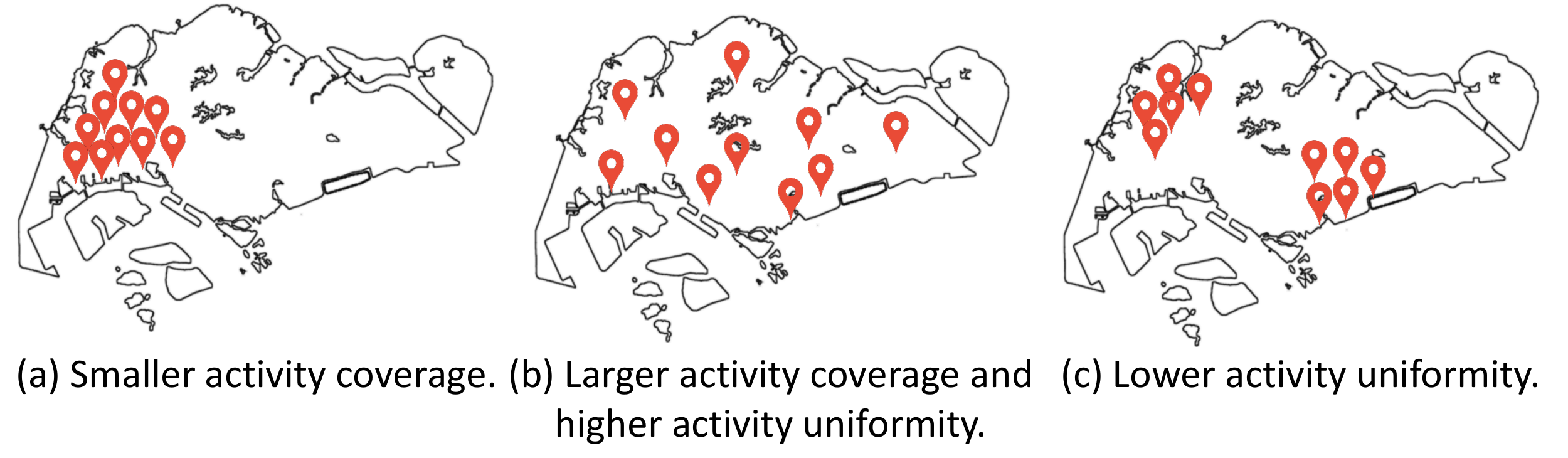}
\caption{Examples of activity coverage and activity uniformity for
different patterns of activity points in reality.}
\label{Fig:activity_patterns}
\end{figure}

Given that privacy is largely a subjective and qualitative notion, this paper proposes to quantify this notion for a set of
location-embedded data submitted by a mobile user, using a metric
called \emph{privacy exposure}. This metric incorporates
\emph{activity coverage} and \emph{activity uniformity}, where the
former refers to the range and the latter refers to the
distribution, of a user's activity hotspots. For an illustrative
example, see \Fig{Fig:activity_patterns}, where the user's
activity coverage in \Fig{Fig:activity_patterns}(a) is much
smaller than in \Fig{Fig:activity_patterns}(b), while her activity
uniformity in \Fig{Fig:activity_patterns}(c) is also much lower
than in \Fig{Fig:activity_patterns}(b). Both cases (a) and (c) are
prone to disclosing activity hotspots of a user.

Secondly, we address the problem of how to minimize privacy
exposure by proposing a privacy-preserving sensing algorithm to
judiciously submit location-embedded data for each mobile user.
Our algorithm (1) exploits \emph{ambient network signatures} which
are composed of the WiFi BSSIDs or cellular base station IDs in
surroundings, in order to avoid exposing activity hotspots, and
(2) cloaks the activity transitions of a user using the {\em
$k$-anonymity} technique. We also devise a learning algorithm to
learn the hotspots so as to differentiate between private and
non-private locations.

We use extensive simulations to investigate how different activity
patterns affect levels of privacy exposure and how our
$k$-anonymity-based technique can reduce privacy exposure. We also
implement our privacy-preserving sensing algorithm in an
Android-based mobile system called WiFi-Scout \cite{WiFi-Scout} to
evaluate the effectiveness of our proposed metric and algorithm.


\section{Related Work}\label{Sec:Related_Work}
Preserving location privacy is a critical issue for mobile LBSs.
The work \cite{Rossi2015_Privacy} studies how location check-in
data in social networks discloses identity of mobile LBS users.
Linear regression models are exploited in
\cite{Ahmadi2010_Regression} to transform data at the client side
before data is submitted so that user privacy can be preserved. In
\cite{Vergara-Laurens2014_privacyQuality}, a hybrid approach is
proposed to preserve location privacy, where the encryption, data
anonymization, and data obfuscation techniques are adaptively
chosen based on the areas of interest. The work
\cite{Agir2014_AdaptiveProtection} provides users with a privacy
guarantee at the lowest utility loss.

Some $k$-anonymity approaches
\cite{Gedik2008_Privacy}\cite{Niu2014_k-anonymity}\cite{Gruteser2003_KAnonymous}
were proposed to achieve privacy preservation in location-based
services. The work \cite{Gedik2008_Privacy} proposes a $k$-anonymity framework with a trusted anonymity server which performs location perturbation on location-embedded queries received from mobile devices before forwarding these queries to the Mobile LBS provider. The work
\cite{Niu2014_k-anonymity} considers an entropy-based privacy metric in the temporal domain
so that the query frequency of the real queried location is
similar to that of other dummy locations. The work
\cite{Gruteser2003_KAnonymous} deals with $k$-anonymous location
information in both spatial and temporal domains when making
queries to LBSs. It proposes an algorithm that chooses a sufficiently
large area and delays the query to make $k$-anonymous queries if necessary.
An entropy-based metric was also used in
\cite{Boutsis2013_PrivacyPreservation} for participatory sensing
applications. It preserves privacy by distributing users' mobility
trajectories among multiple databases, and balancing the
frequencies of locations in each individual database. However,
there still lacks a metric for quantifying privacy exposure
particularly in the spatial domain.

Compared with existing research efforts, the key contributions of this
paper are summarized as follows. First, our work is the first to
formally quantify privacy exposure in the spatial domain for a
given set of activity points of an arbitrary user. The metric is
generally defined such that it can be applied to various mobile
LBSs. Second, we formally define the privacy exposure problem, and
propose an algorithm as a solution. Finally, we evaluate our
proposed metric and algorithm via both simulations and real-world
experiments, which confirm their validity and effectiveness.

\section{Problem Definition}\label{Sec:System Model}

\subsection{Countermeasures to Privacy Threats} 
Multi-modal built-in sensors on smartphones have urged many mobile
LBSs which highly rely on location-embedded information of users.
To explain the privacy threats and how we can protect user privacy
in mobile LBSs, we consider a WiFi advisory system,
\emph{WiFi-Scout} \cite{Wu2015_ICC}, as an example.
WiFi-Scout relies on crowdsourced WiFi-quality data to help
smartphone users to find nearby WiFi access points (APs) with good
quality. It provides two operating modes, \emph{query mode} and
\emph{crowdsensing mode}, for mobile users. The query mode allows
a user to search for WiFi APs with high quality ranking in the
proximity of the user's current location, so that the user can
plan the next move or choose a good meeting point with friends.
The crowdsensing mode allows a user who already connects to a WiFi
AP to rate the AP, according to the actual experience of using the
AP, through the user's smartphone. Both modes require users'
location information: the query mode needs to know the location of
the user in order to search for nearby APs, and the crowdsensing
mode also needs to obtain the user's location together with the
associated WiFi AP quality information.

However, if the back-end system is not trustworthy or properly
designed, the sensitive information of users may be disclosed to
adversaries. Specifically, we consider two types of privacy
threats in this work: \emph{activity hotspot disclosure} and
\emph{activity transition disclosure}. A user's activity hotspots
are the locations or places she stays for a long time or
frequently visits, such as home, office, and her favorite cafe. A
user's activity transition is the moving sequence that connects
all the user's activity hotspots.

Conventionally, the countermeasures to privacy threats can be
classified into three categories: (1) anonymization, (2)
obfuscation, and (3) encryption. Anonymization is to
``disidentify'' a user by, for example, reporting additional
locations (typically of other users) together with the user's
location. Obfuscation is to report a ``modified'' version instead
of the real location of a user. Encryption exploits cryptographic
methods to conduct security key encryption and decryption between
front-end clients and back-end servers. The first and second
countermeasures are performed at front-end clients whereas the
third relies on collaboration between front-end clients and
back-end servers. In this paper, we combine anonymization and
obfuscation approaches so that front-end sensing clients have full
control over the reported data in order to protect user privacy.

\begin{figure}
\centering
\includegraphics[width=\columnwidth]{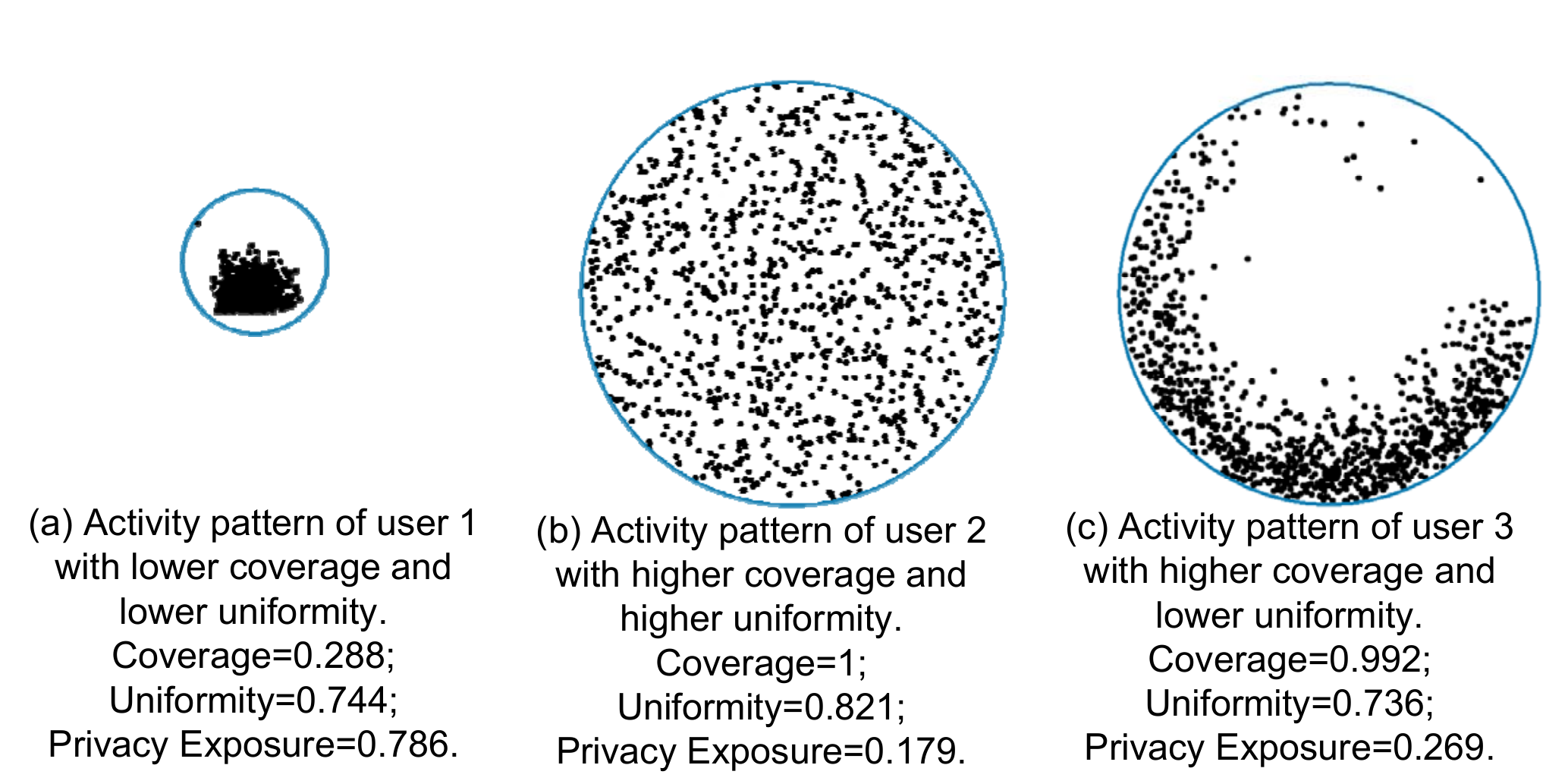}
\caption{Privacy exposure for three users each with 1000 activity points (marked as black dots), where the blue circle is the smallest disc enclosing the 1000 activity points.}
\label{Fig:privacy-exposure}
\end{figure}

\subsection{The Privacy Exposure Problem} \label{subsec:PrivacyExposureProblem}
This work sets out to answer the following two questions:
(1) how to quantify \emph{privacy exposure} for a given set of
data submitted by a certain user and (2) how to effectively reduce
privacy exposure. Note that the ``data" here is location-tagged.
Assume that each user has a unique user account in the system
(this could be for the rewarding purpose in order to encourage user
participation and improve trustworthiness
\cite{Foursquare}\cite{sew14}). Let $\Omega$ denote the set of
reports in the WiFi advisory system. Let $r_i=(w_i, l_i, a_i, s_i,
q_i, f_i)$ denote the $i$-th report, where $w_i$ is the ID of the
WiFi access point that the smartphone is associated with, $l_i$ is
the location of the smartphone, $a_i$ is the positioning accuracy
(e.g., obtained from Google localization API), $s_i$ is the signal
strength received by the smartphone, $q_i$ is the link quality
(i.e., link speed) observed by the smartphone, and $f_i$ is the
set of ambient visible WiFi access points observed by the
smartphone. Note that each report is submitted to the system when
the smartphone is under either the query mode or the crowdsensing
mode.

Given a set of reports $\{r_1, r_2, \ldots,r_n \}$ submitted by a
particular user, we define \emph{privacy exposure} based on the
following two concepts: (1) \emph{activity coverage} (or
``coverage" for short) and (2) \emph{activity uniformity} (or
``uniformity" for short). The former refers to the location range
of the user's all activities, while the latter refers to how
uniform the user moves within the above range.
\begin{dfn}
Given a set of reports $\{r_1, r_2, \ldots,r_n \}$ submitted by a
particular user, the \emph{activity coverage} is defined as
\[
\lambda(\{r_1, r_2, \ldots,r_n \})=\frac{D}{D_{max}},
\]
where $D$ is the diameter of the smallest circle covering all the
reports and $D_{max}$ is a normalizing constant no smaller than $D$,
such that $0\leq \lambda(\{r_1, r_2, \ldots,r_n \})\leq 1$.
\end{dfn}

Finding the value of $D$ for a given set of reports is a classic
smallest enclosing disc problem \cite{ComputationalGeometryBook},
to which there are many approximation algorithms as solutions.
However, the computational complexity of these approximation algorithms
does not suit smartphones. Therefore, for a practical
implementation, we approximate this diameter as
$D=\min\{D_{max}, 2\times \max\{d(\frac{\sum_{i=1}^n l_i}{n},
l_j)|j=1,2,\ldots, n\} \}$, where $d(\frac{\sum_{i=1}^n l_i}{n},
l_j)$ is the distance between the center of gravity of all the
reports' locations and the location $l_j$.
\Fig{Fig:privacy-exposure} gives an example of 1000 activity
points for 3 different users. \Fig{Fig:privacy-exposure}(a) shows
that the activity pattern of user 1 has lower coverage compared to
the user in \Fig{Fig:privacy-exposure}(b).
\begin{dfn}
Given a set of reports $\{r_1, r_2, \ldots,r_n \}$ submitted by a
particular user, the \emph{activity uniformity} is defined as
\[
U(\{r_1, r_2, \ldots,r_n \})= \frac{[\sum_{i\neq j} d(l_i,
l_j)]^2}{{n\choose 2} \cdot \sum_{i\neq j} d^2(l_i, l_j)}.
\]
\end{dfn}
The uniformity $U$ is the equilibrium level of all the distances between
each pair of reports and is defined using Jain's fairness index
\cite{Jain91_fairnessIndex}; 
its value $0 \leq U(\{r_1, r_2, \ldots,r_n \})\leq 1 $.

Now we can define define privacy exposure based on the coverage
and uniformity defined above.

\begin{dfn}
Given a set of reports $\{r_1, r_2, \ldots,r_n \}$ submitted by a
particular user, the \emph{privacy exposure} of the user is
defined as
\begin{multline}\label{eq:privacy_expo}
\Psi(\{r_1, r_2, \ldots,r_n \})= \\1 - \lambda(\{r_1, r_2, \ldots,r_n \})
\cdot U(\{r_1, r_2, \ldots,r_n \}).
\end{multline}
\end{dfn}
Therefore, a lower coverage or a lower uniformity results in a
higher privacy exposure, while a lower exposure is preferred.
Also, $0 \leq \Psi(\{r_1, r_2, \ldots,r_n \})\leq 1$.
\Fig{Fig:privacy-exposure}(b) shows that the activity pattern of
user 2 has higher uniformity compared with the user in
\Fig{Fig:privacy-exposure}(c).

Thus, we have answered the first question, i.e., how to quantify
the privacy exposure for a given set of reports submitted by a
particular user. To answer the second question, i.e., how to
reduce privacy exposure when the user needs to make location-based
queries or submit location-tagged reports, we define the
\emph{privacy exposure problem}, taking a divide-and-conquer
approach.

\begin{dfn}
Given a crowdsourced dataset $\Omega$ from all the users and the
set of reports $\{r_1, r_2, \ldots,r_n \}$ submitted by a
particular user, the \emph{privacy exposure problem} is to
determine, for a new report $r$ sensed by this user: (1) whether
it is necessary to submit the new report $r$ and (2) how to cloak
this report to minimize privacy exposure if the submission is
necessary.
\end{dfn}

\section{A Privacy-preserving Sensing Algorithm} \label{Sec:Algo}

In this section, we design a dual-mode sensing algorithm to solve
the above-defined privacy exposure problem. We differentiate user
locations (activity hotspots) between \emph{private places} and
{\em non-private places}. The private places refer to a user's
``long-term stay'' places, such as home and office. The
non-private places refer to ``short-term stay'' points of interest
(PoIs) that are frequently visited by the user, such as her
favorite cafes. Both places should not be directly submitted
without control, which may otherwise lead to lower coverage and
lower uniformity, and hence higher privacy exposure.

\begin{figure}
\centering
\includegraphics[width=0.8\columnwidth]{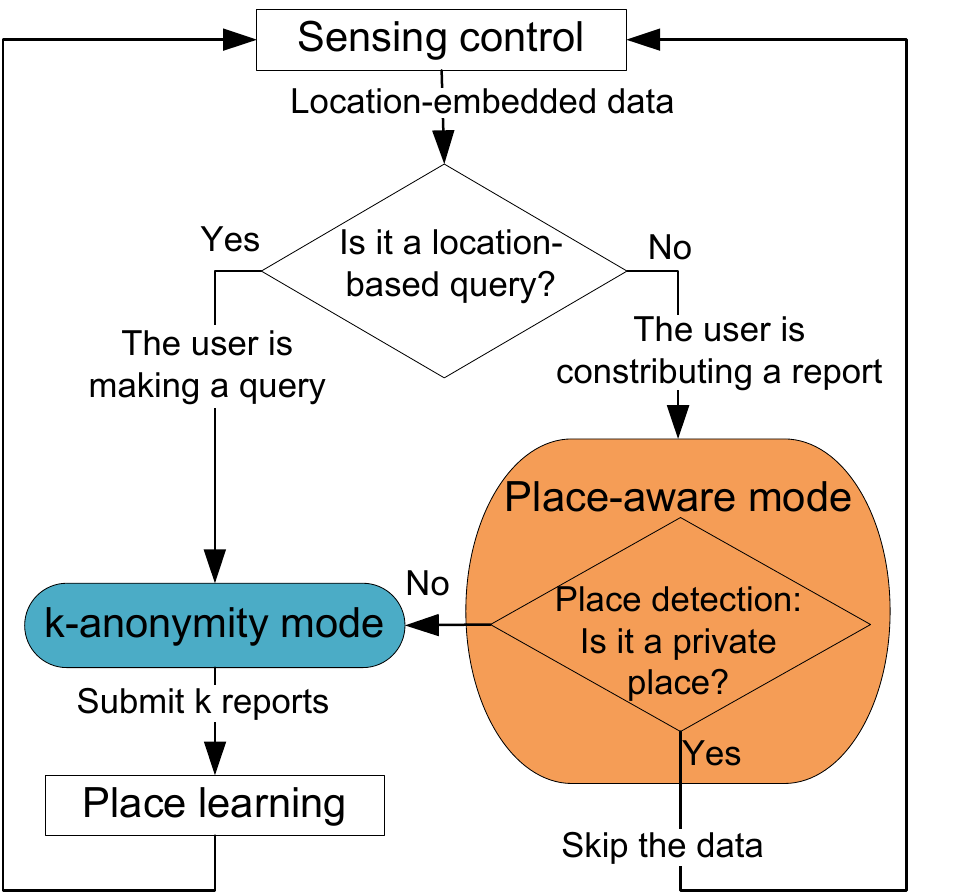}
\caption{Workflow of our privacy-preserving sensing algorithm.}
\label{Fig:dual-mode-algo}
\end{figure}

Our dual-mode sensing algorithm is illustrated in
\Fig{Fig:dual-mode-algo}. Each smartphone operates in one of the
two modes, \emph{place-aware mode} and \emph{$k$-anonymity mode},
when holding a location-tagged report subject to possible
submission. If the user is in a private place, the smartphone will
operate in the place-aware mode which will skip all the
location-tagged data. Otherwise, the smartphone operates in the
$k$-anonymity mode which will choose $k-1$ additional reports to
submit together with the original report.

\subsection{Place Learning and Detection in Place-aware Mode}
We exploit the network fingerprint-based place learning and
detection \cite{Wu2014_UrbanMobilitySense} to determine if a
smartphone should submit its sensed location-embedded data. Each
smartphone maintains the WiFi fingerprints of the user's private
place profile $\mathcal{P}$. Initially, $\mathcal{P}=\emptyset$.
When the user's smartphone senses location-embedded data in an
unknown place at time $t$, it starts to learn the WiFi
fingerprints of this unknown place, denoted by $p_i=q(t)$, where
$q(t)=\{w_1, w_2,\ldots, w_k\}$ includes all the visible WiFi BSSIDs
at time $t$. The $p_i$ will be updated by $p_i=p_i\cap q(t+1)$ at
time $t+1$ incrementally if $S(p_i, q(t+1))\geq \Delta$, where
$S(p_i, q(t+1))=\frac{|q(t+1)\bigcap p_i|}{|p_i|}\in[0,1]$ is the
similarity between the sensed location and the current WiFi
fingerprints of the private place $p_i$, and $\Delta$ is a
configurable threshold. Otherwise, if $S(p_i, q(t+1))<\Delta$, it
gives up $p_i$ as a private place since the user is detected to
have left $p_i$. If $p_i \neq \emptyset$ till $t+k$ where
$k>\Delta_L$ and $\Delta_L$ is a configurable threshold,
$\mathcal{P}$ is updated as $\mathcal{P}=\mathcal{P}\cup \{p_i\}$,
and the above learning process continues. 
Furthermore, for each location sensed by the smartphone at time $t$,
the algorithm will also conduct the place detection by calculating
the similarity $S(q(t),p_j)$ for each $p_j\in \mathcal P$.
If there exists a $p_j\in \mathcal P$ with
$S(q(t),p_j)\geq \Delta$, then the user is in the private place
$p_j$ and $q(t)$ is discarded.

The aforementioned scheme of place learning and detection is for
submitting crowdsensed data as indicated in the right-hand portion of
\Fig{Fig:dual-mode-algo}. When submitting a query, on the other
hand, it will not enter the place-aware mode and the query will be
submitted anyway. However, the $k$-anonymity mode will be
activated to protect privacy as explained next.

\subsection{Data Anonymization in $k$-anonymity Mode}
In the $k$-anonymity mode, we exploit the $k$-anonymity technique
\cite{Niu2014_k-anonymity} to choose $k-1$ additional reports to
submit together with the real report in order to reduce privacy
exposure. Our algorithm consists of two phases:
\emph{anonymization phase} and \emph{obfuscation phase}. The
anonymization phase determines the extra $k-1$ reports, and the
obfuscation phase modifies the original report to avoid giving
away the user's exact location.
\begin{enumerate}
\item{\it Anonymization phase}: Let $\Phi$ denote the $k$-anonymity
set which will be submitted by the smartphone. When the smartphone
senses a new report denoted by $r_j=(w_j, l_j, a_j, s_j, q_j,
f_j)$, the $k$-anonymity set is initialized as $\Phi=\{r_j\}$. Then
our algorithm will update $\Phi$ iteratively until $|\Phi|=k$, as
follows.

\begin{enumerate}
\item Each smartphone will maintain, or retrieve from server, two subsets
of reports denoted by $C$ and $\overline{C}=\Omega-C$,
respectively, where $C$ is the set of reports already submitted by
the current user's smartphone and $\overline{C}$ is the set of
reports submitted by other users in the system. Note that since
$\Omega$ is a crowdsourced dataset contributed by all of users in
the mobile LBS, each user's smartphone can only differentiate $C$
from $\overline{C}$ but cannot know the linkage between other
users and their reports.

\item If $\overline{C}\neq\phi$, the smartphone will select a
report
\[ r_p = \argmin_{r\in \overline{C}} \Psi(\Phi \bigcup \{r \}) \]
where $\Psi(\cdot)$ is defined in \eqref{eq:privacy_expo}.

\item Otherwise, select a report $r_q \in C$ whose location appears
in $C$ for the minimum number of times.

\item Update $\Phi=\Phi \bigcup \{r_p \}$.

\item Repeat (b)--(d) until $|\Phi|=k$.

\end{enumerate}

\item{\it Obfuscation phase:} Given the real report $r_j=(w_j, l_j, a_j,
s_j, q_j, f_j)$, find $l_{PoI}$ (e.g., via the Google Place API \cite{GooglePlaceAPI})
which is the nearest point of interest to $l_j$, and update replace $l_j=l_{PoI}$.

\end{enumerate}

\section{Simulation}\label{Sec:Simulations}

\subsection{Simulation Setup}

We conduct extensive simulations to examine our proposed metric of
privacy exposure defined in \Sec{Sec:System Model} and to evaluate
our algorithm outlined in \Sec{Sec:Algo}. First, we simulate
different activity patterns to see how activity patterns affect
the values of coverage, uniformity, and privacy exposure. Then, we
compare our $k$-anonymity mode against a \emph{random
$k$-anonymity} (``random" for short) and a \emph{naive} (``naive"
for short) sensing algorithm to investigate the performance. The
random $k$-anonymity sensing algorithm chooses the $k-1$ redundant
reports randomly, and the naive sensing algorithm submits reports
directly without any cloaking mechanism.

\begin{figure}
\centering
\includegraphics[width=0.85\columnwidth]{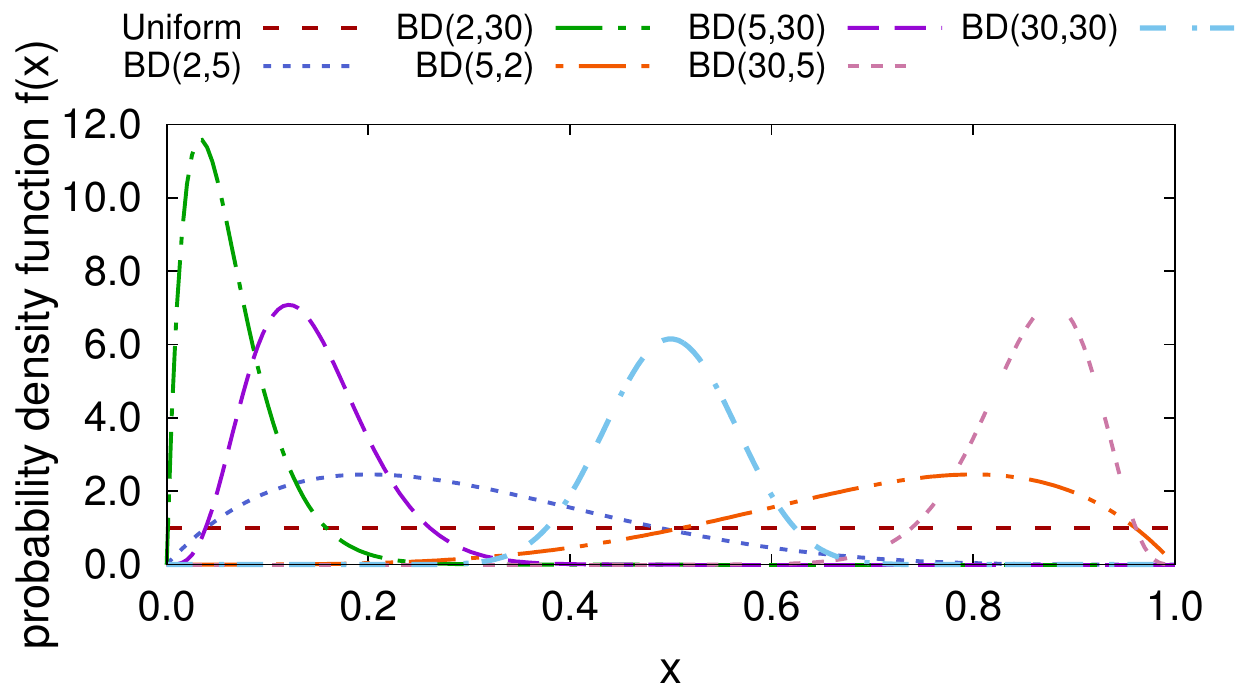}
\caption{Uniform and beta distributions for generating activity patterns.}
\label{Fig:ProbDistri}
\end{figure}

In our simulations, the experimental field is a circle with
diameter $D_{max} = 500$ meters. The activity patterns of users
are generated by the uniform and beta distributions in
\Fig{Fig:ProbDistri}, where $BD(\alpha,\beta)$ denotes the beta
distribution with shape parameters of $\alpha$ and $\beta$. We
consider the polar coordinate system to generate activity points
within the experimental field, where each activity point is
represented as a pair of $(\gamma,\theta)$. Here,
$\gamma=\sqrt{\rho} \times \frac{D_{max}}{2}$ is the distance from
the activity point to the center of the experimental field, $\rho
\in [0,1]$ is a random number generated by the probability
distributions in \Fig{Fig:ProbDistri}, and $\theta = \rho \times
2\pi$ is the polar angle of the activity point. Thus, a smaller
value of $\alpha$ and a larger value of $\beta$ will generate
activity points close to the center of the experimental field and
span a narrow angle in the experimental field. In this way, we can
generate non-uniform and dense cases of activity points to study
the proposed metrics. All the simulation results are averaged over
100 runs.

\subsection{Simulation Results}

\begin{figure*}[!t]
\centering
\begin{tabular}{ccc}
\includegraphics[width=0.32\linewidth]{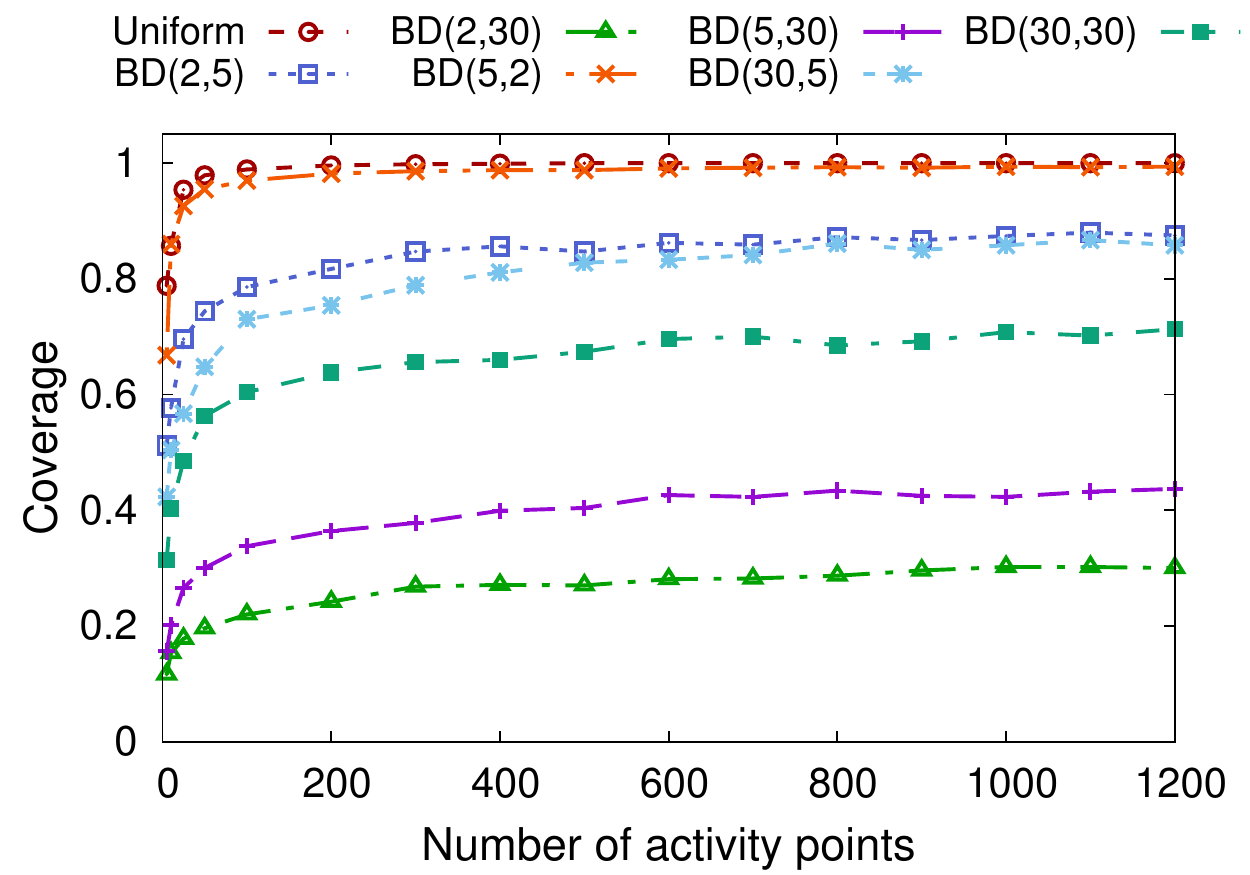} &
\includegraphics[width=0.32\linewidth]{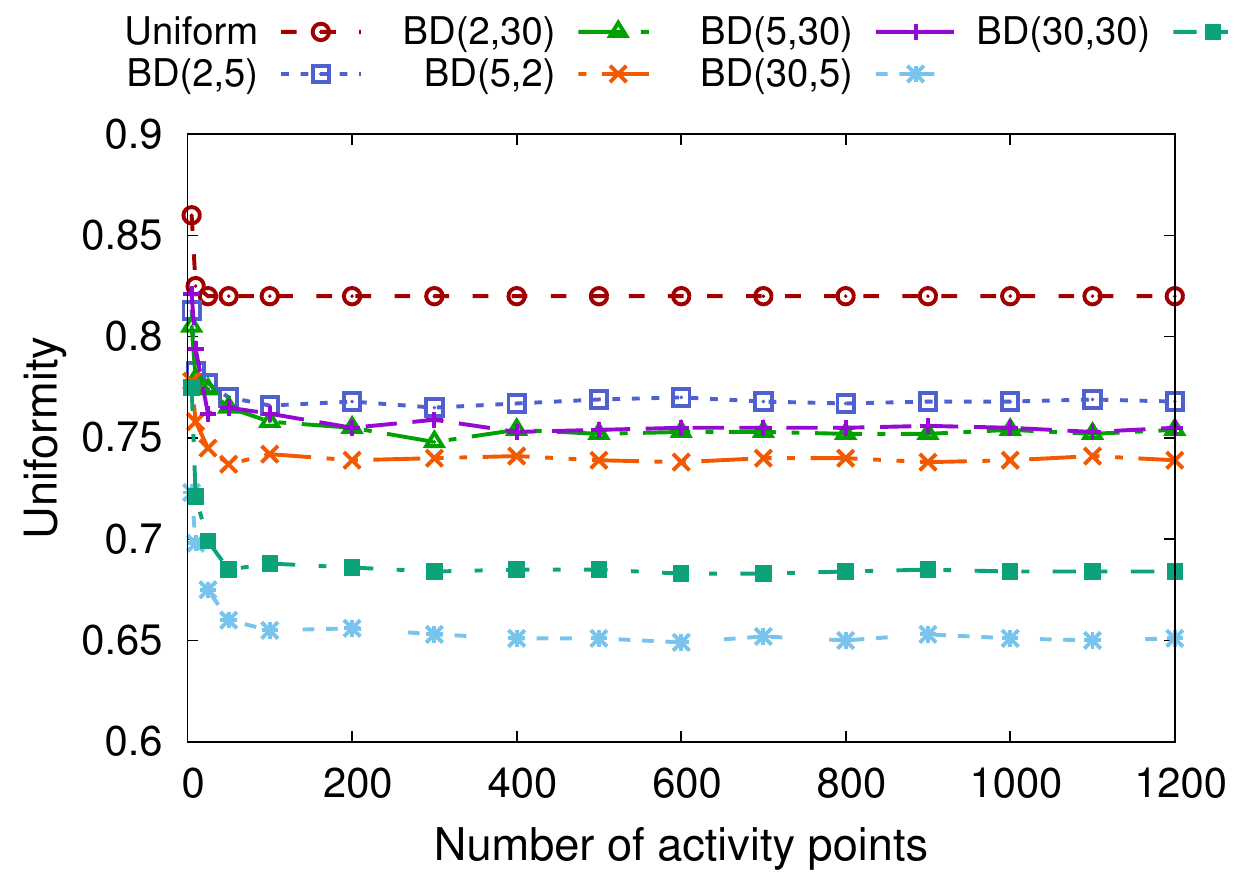} &
\includegraphics[width=0.32\linewidth]{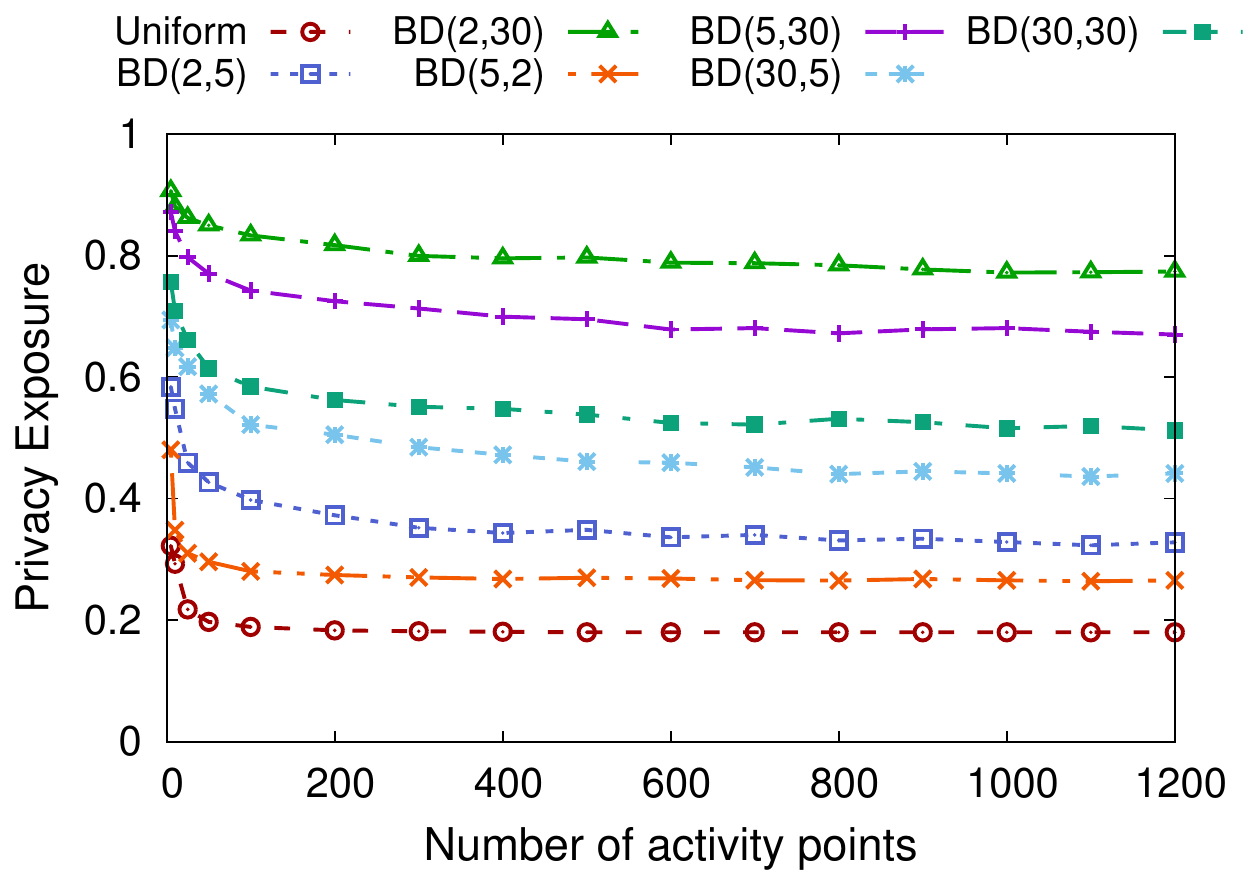}\\
(a) & (b) & (c)\\
\end{tabular}
\caption{Comparison of (a) coverage, (b) uniformity, and (c)
privacy exposure for different activity patterns.}
\label{Fig:ExpComparison}
\end{figure*}

In the first set of simulations, we vary the number of activity
points of a single user from 0 to 1200 to study the coverage,
uniformity, and privacy exposure. \Fig{Fig:ExpComparison}(a)
presents the simulation results. As it can be seen, the coverage
of the activity points generated by the uniform distribution and
$BD(5,2)$ distribution are higher than others because their radii
of the smallest enclosing circles are close to the radius of the
experimental field. On the other hand, $BD(2,30)$ results in lower
coverage because its activity points are located at an extremely
small area around the center of the experimental field. For
uniformity, as shown in \Fig{Fig:ExpComparison}(b), $BD(30,5)$ and
$BD(30,30)$ have lower uniformity because both distributions
generate activity points within a ring-belt area along the
boundary of the experimental field. Finally, for privacy exposure
as shown in \Fig{Fig:ExpComparison}(c), we see that the uniform
distribution has the lowest privacy exposure while the $BD(2,30)$
distribution has the highest. This implies that coverage weighs
more than uniformity in privacy exposure, which conforms to our
intuition. Thus, this set of simulations demonstrate that our
proposed privacy exposure is a sensible metric to evaluate user
privacy in the spatial domain as users who have a larger activity
coverage, and more uniform activity points will have a better
(lower) privacy exposure.

An additional observation is that the privacy exposure
does not improve when the number of activity points is
larger than $200$. This is because the activity points generated
in the beginning are ``representative'' points that largely determine
the coverage, uniformity, and privacy exposure.

\begin{figure*}[!t]
\centering
\begin{tabular}{ccc}
\includegraphics[width=0.32\linewidth]{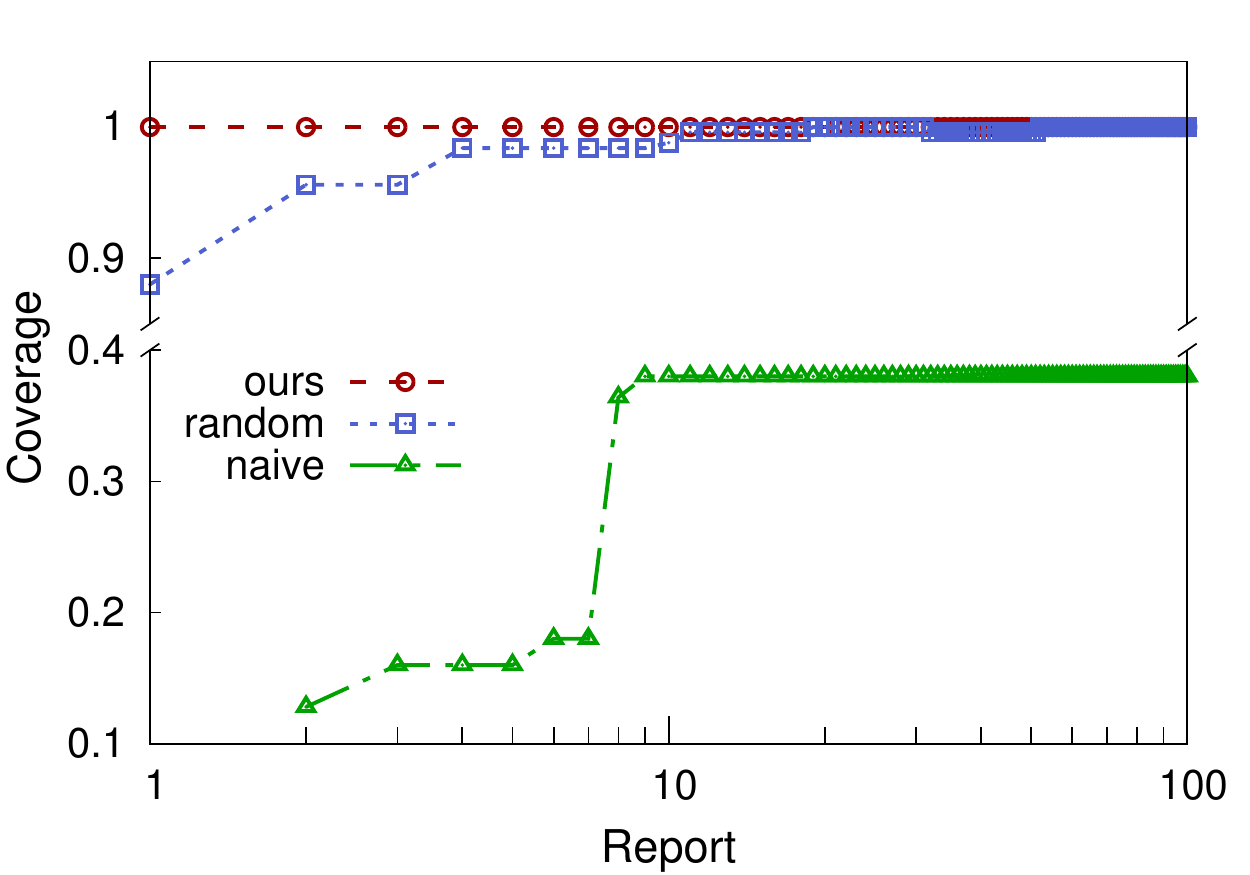} &
\includegraphics[width=0.32\linewidth]{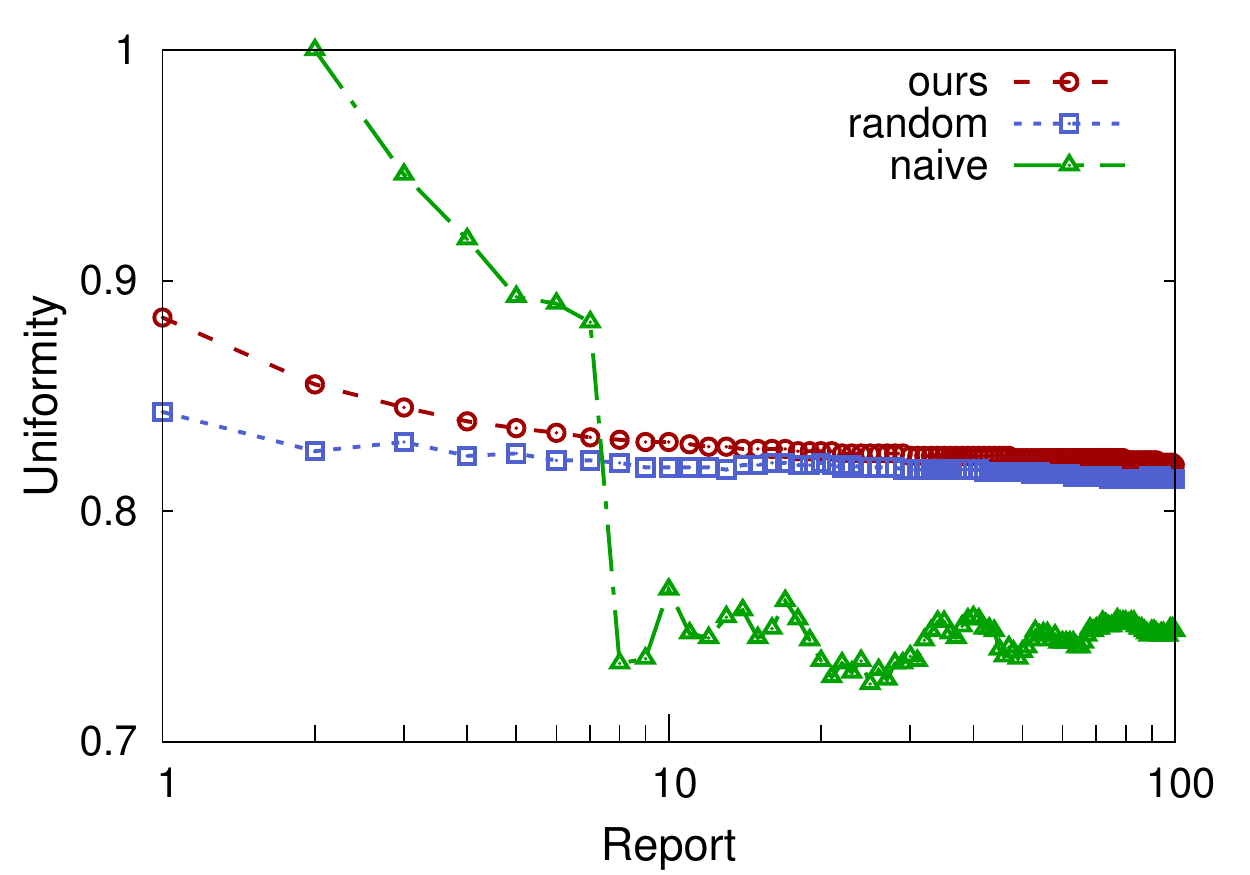}&
\includegraphics[width=0.32\linewidth]{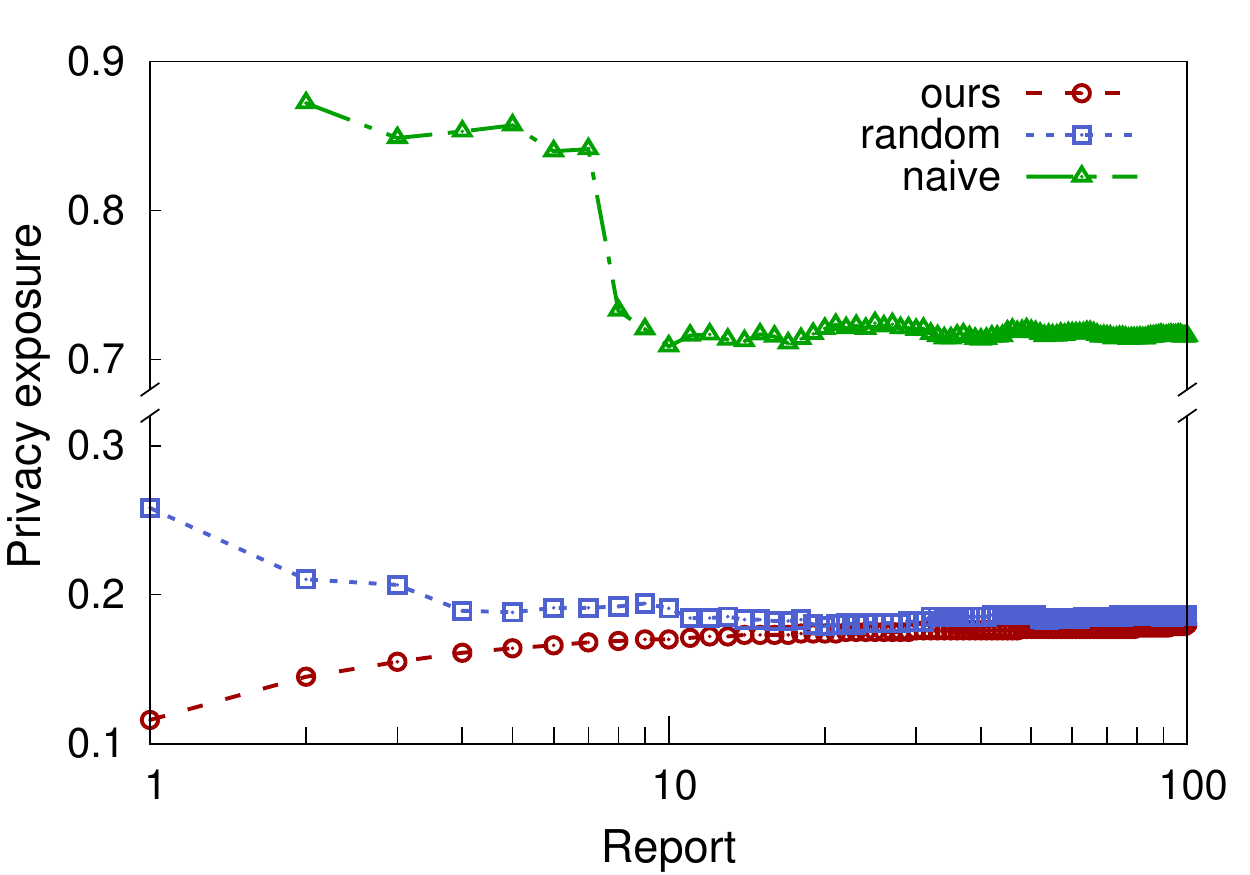} \\
(a) & (b) & (c)\\

\includegraphics[width=0.32\linewidth]{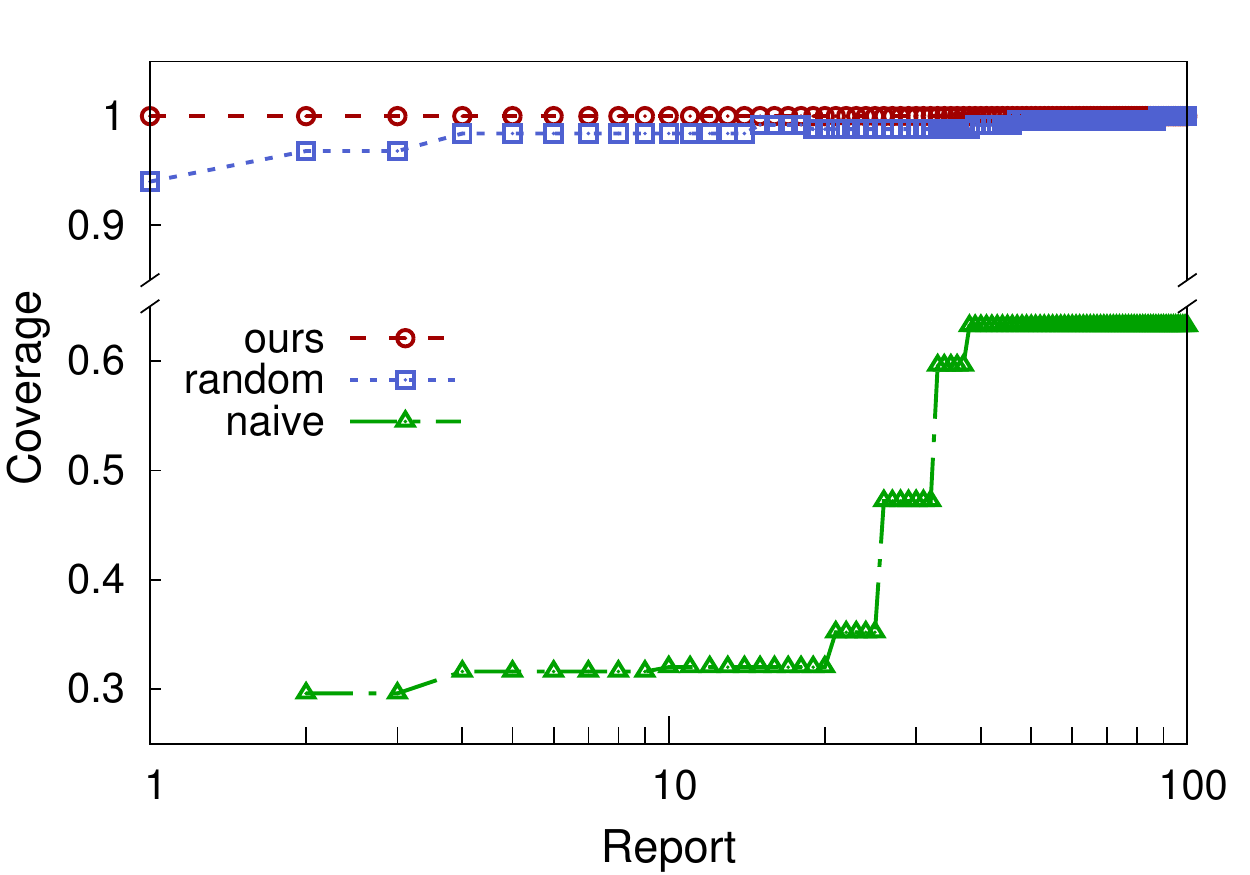} &
\includegraphics[width=0.32\linewidth]{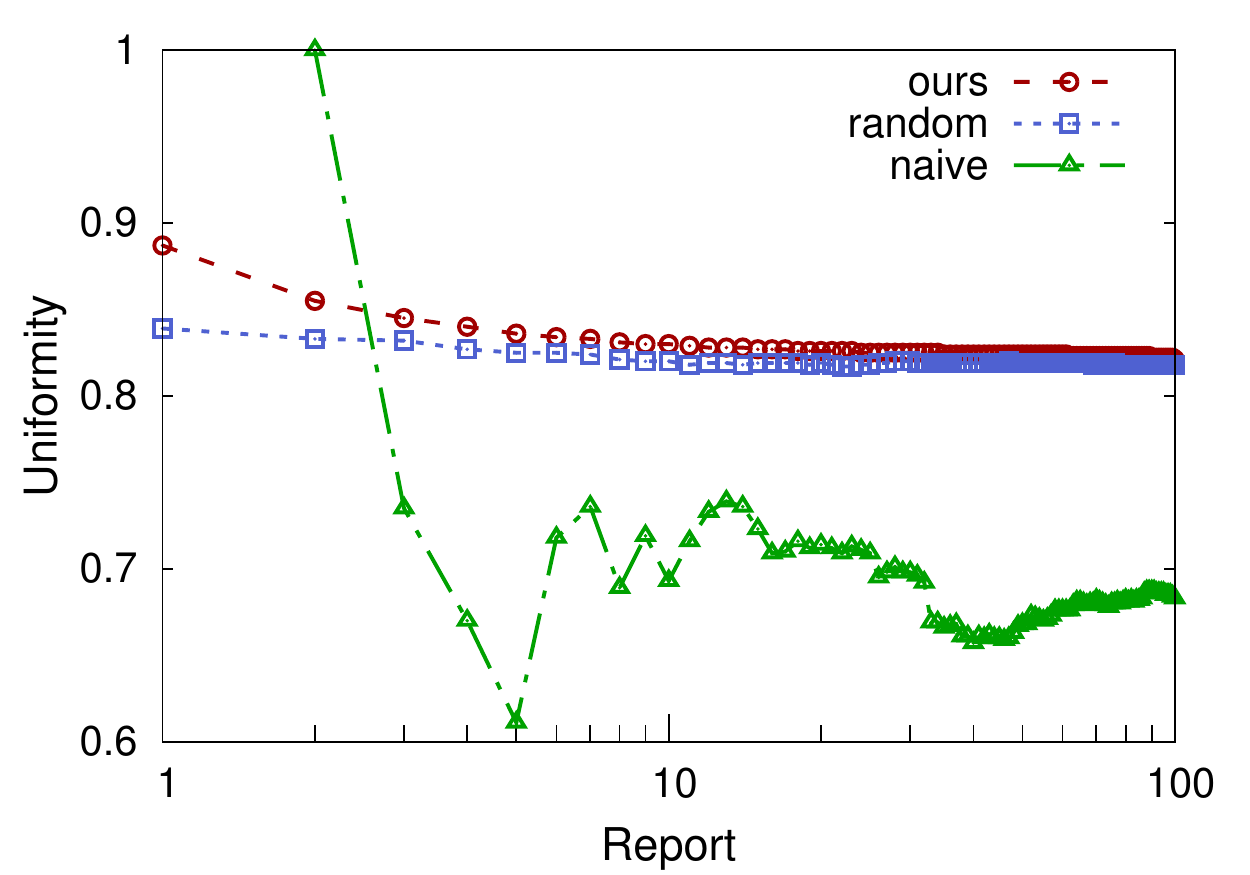}&
\includegraphics[width=0.32\linewidth]{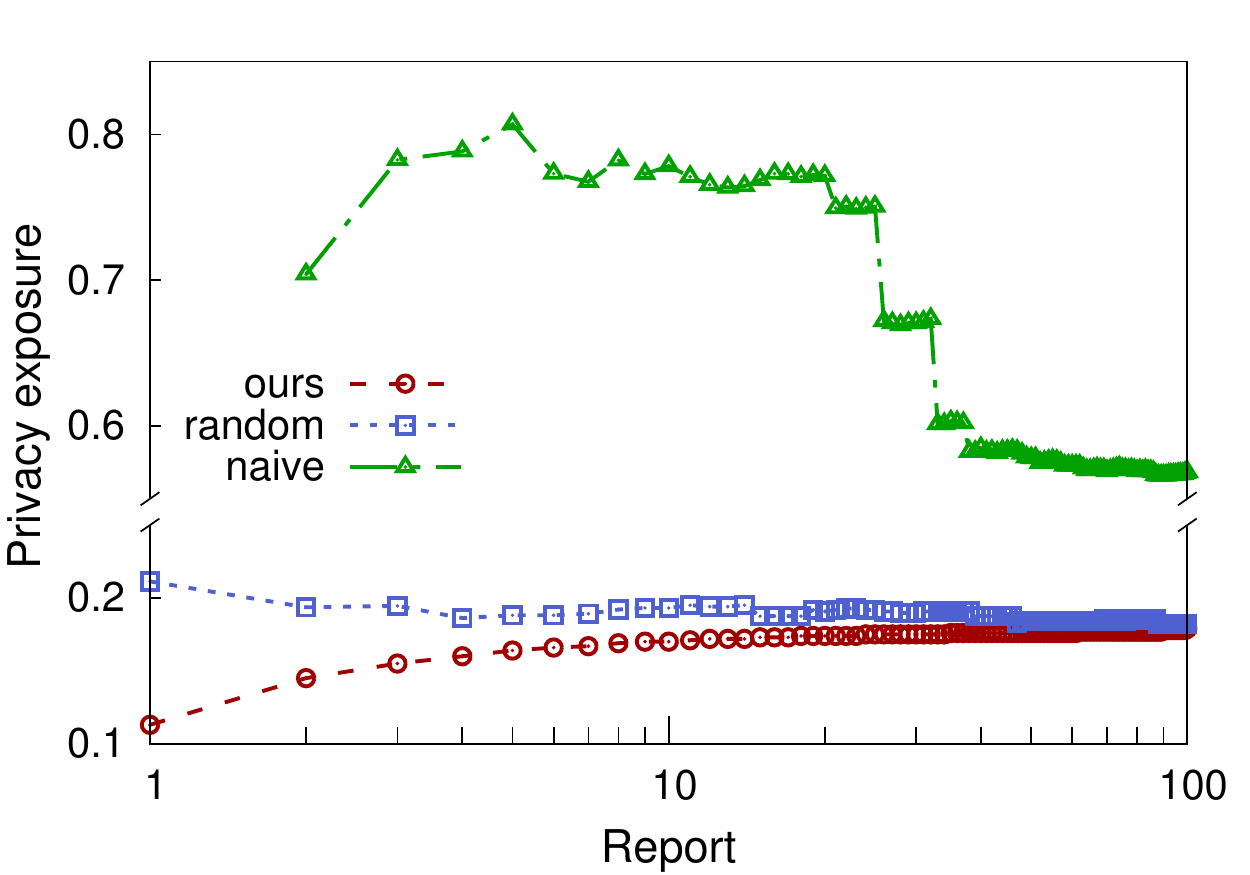} \\
(d) & (e) & (f)\\
\end{tabular}
\caption{Comparison of coverage, uniformity, and privacy exposure for different
algorithms. In our $k$-anonymity algorithm and {\it random}, $k=10$.
The activity points in (a)-(c) are generated by $BD(5,30)$ and
the activity points in (d)-(f) are generated by $BD(30,30)$.} \label{Fig:AgloSimResults}
\end{figure*}

In the second set of simulations, we compare our $k$-anonymity
mode against the random and the naive sensing algorithms. We
uniformly generate $\Omega$ with $1000$ reports in the
experimental field. In the simulation, we consider a single user
who will sequentially submit $100$ real reports using our
$k$-anonymity mode with $k=10$, where the activity points of the
user are generated by $BD(5,30)$ and $BD(30,30)$ distributions,
respectively. We choose these two distributions because $BD(5,30)$
will generate activity points within a small area closer to the
center of the experimental field while $BD(30,30)$ will generate
activity points within a ring-belt area along the boundary of the
experimental field. \Fig{Fig:AgloSimResults} gives the result,
where for each submitted new report, we trace the changes in
coverage, uniformity, and privacy exposure.
\Figss{Fig:AgloSimResults}(a)-(b) and
\Figss{Fig:AgloSimResults}(d)-(e) show that our algorithm leads to
higher coverage and higher uniformity than the random and the
naive sensing algorithms. This is because of the judicious choice
of the $k-1$ additional data in our algorithm to achieve the
lowest privacy exposure. In \Fig{Fig:AgloSimResults}(c) and
\Fig{Fig:AgloSimResults}(f), our algorithm improves privacy
exposure by $74.9\%$ and $68.5\%$, respectively, as compared to
the naive sensing algorithm at the end of the simulation. On the
other hand, the privacy exposure of the random sensing algorithm
is closer to the result of ours when the simulations terminate.
This is because when the number of submitted reports increases,
the random sensing algorithm will gradually improve the coverage
and approach the performance of our algorithm. However, it is
important to note that our algorithm has optimized both coverage
and uniformity even in the very beginning when there are only few
user activity points. Thus, our algorithm allows users to submit
location-tagged reports without significantly exposing privacy for
any user activity patterns.

\begin{figure*}[!t]
\centering
\begin{tabular}{cccc}
\includegraphics[width=0.23\linewidth]{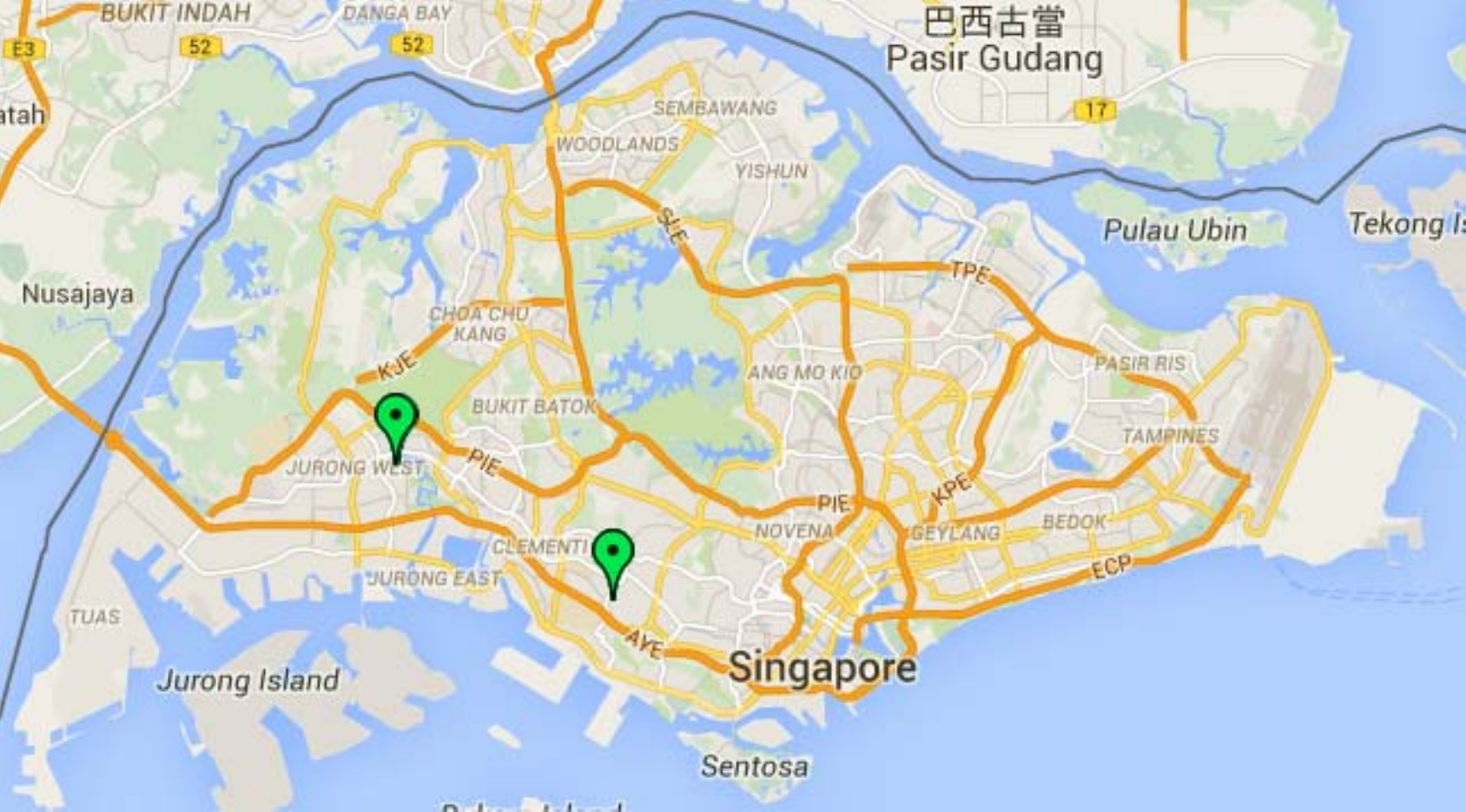} &
\includegraphics[width=0.23\linewidth]{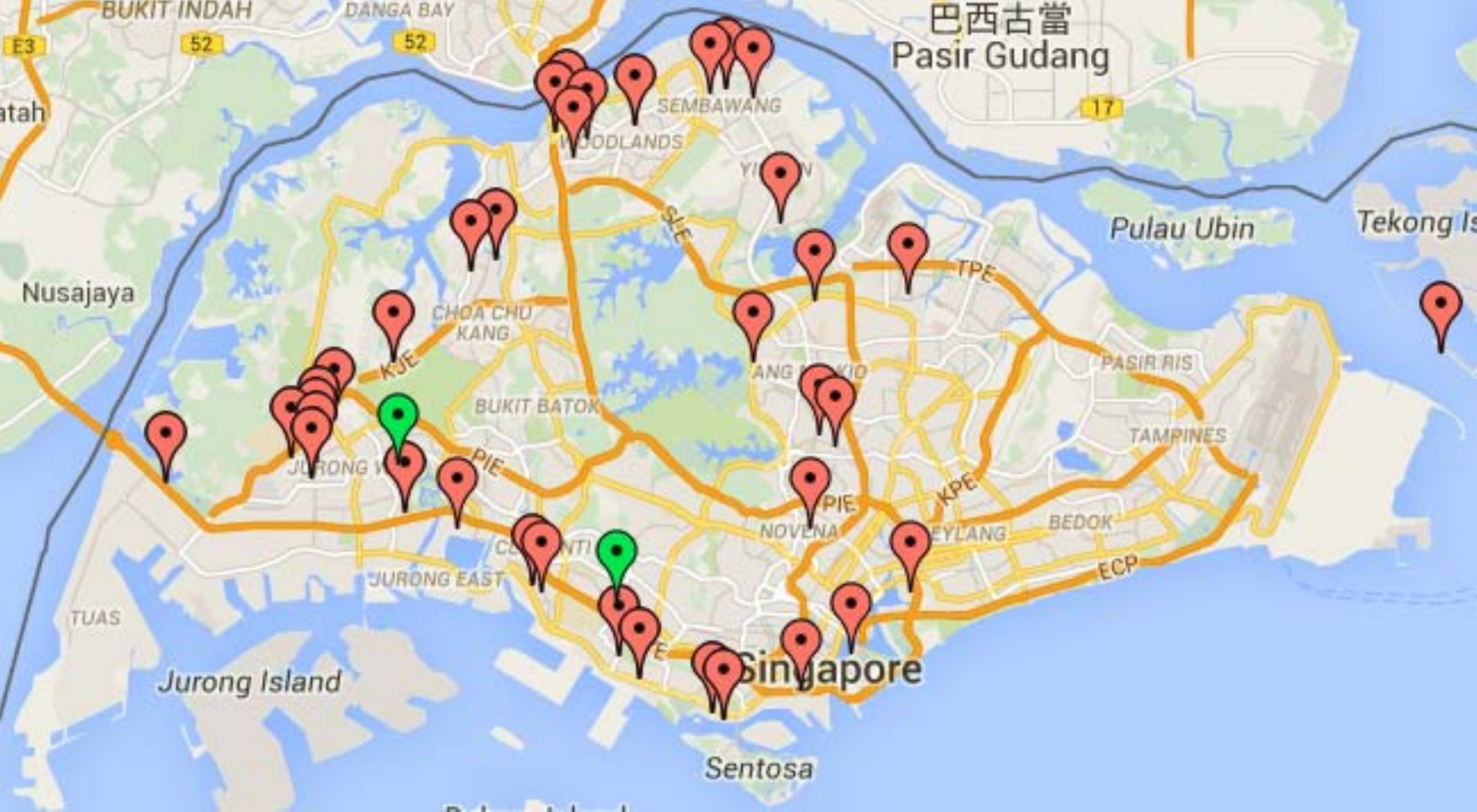} &
\includegraphics[width=0.23\linewidth]{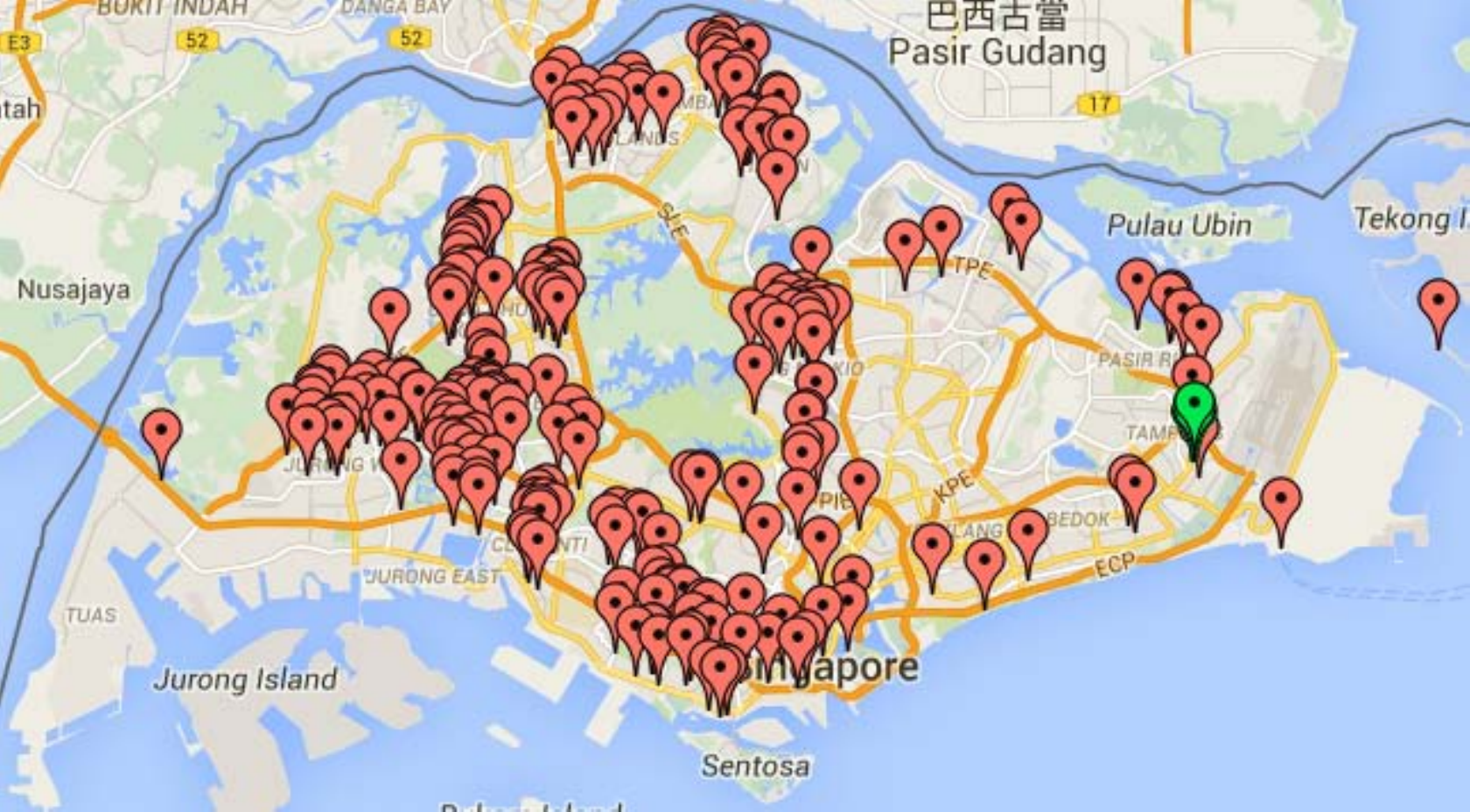} &
\includegraphics[width=0.23\linewidth]{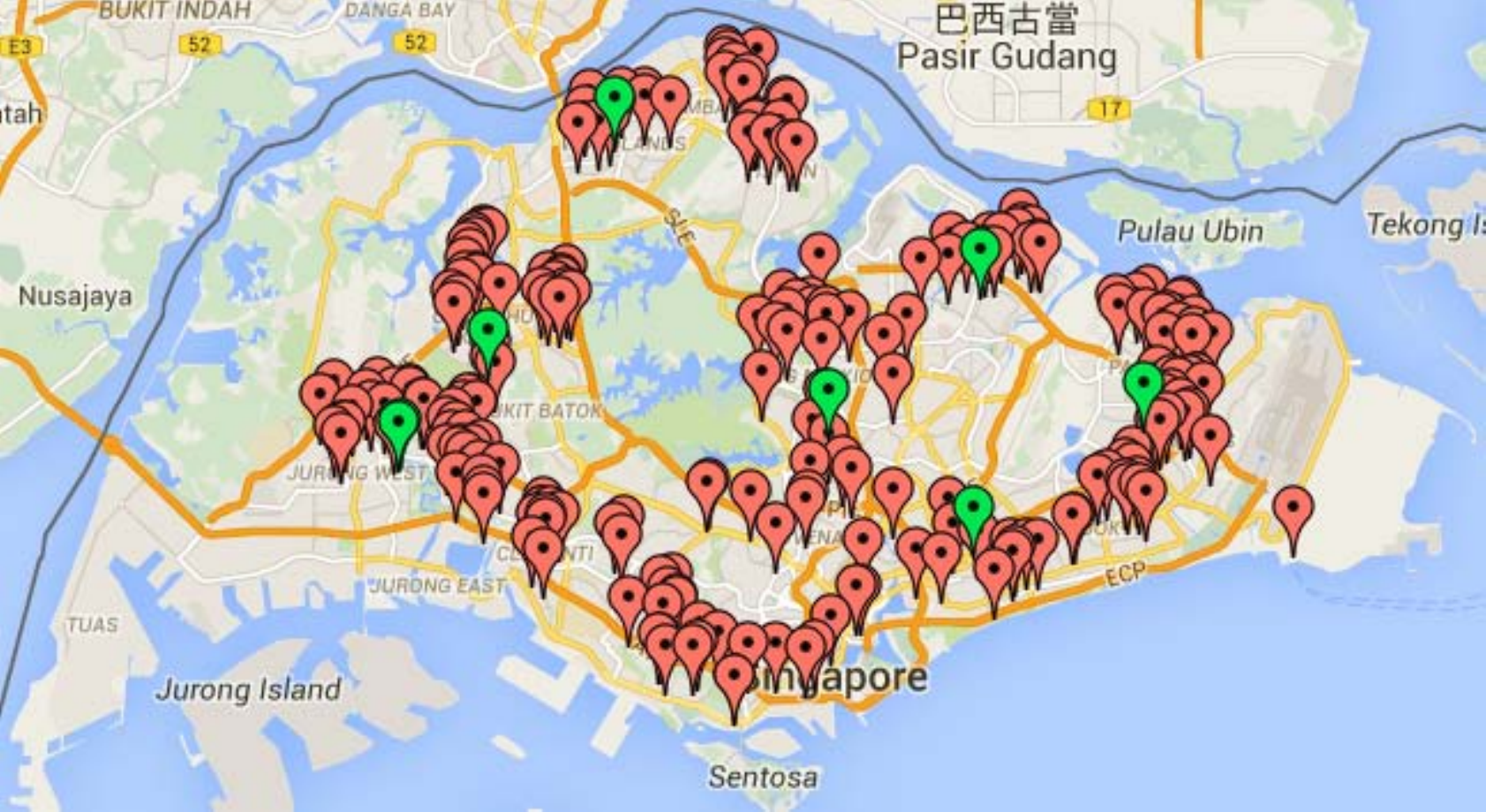} \\
(a) & (b) & (c) & (d)
\end{tabular}
\caption{Experimental results: (a) 34 real reports submitted by
user 1's smartphone 1 running the {\em naive} sensing algorithm, where
coverage=$0.184$, uniformity=$0.515$, and privacy
exposure=$0.905$; (b) 60 reports inclusive of 2 real reports ($k=30$)
submitted by user 1's smartphone 2 running our sensing algorithm,
where coverage=$1.0$, uniformity=$0.745$, and privacy
exposure=$0.255$; (c) $240$ reports inclusive of 8 real reports
submitted by user 2's smartphone running our sensing algorithm, where
coverage=$1.0$, uniformity=$0.759$, and privacy
exposure=$0.241$; (d) $240$ reports inclusive of 8 real reports
submitted by user 3's smartphone running our sensing algorithm, where
coverage=$0.729$, uniformity=$0.787$, and privacy
exposure=$0.426$.} \label{Fig:ExpResults}
\end{figure*}

\section{Implementation} \label{Sec:Implementation}

\subsection{Setup}
We have implemented our dual-mode sensing algorithm in WiFi-Scout
\cite{Wu2015_ICC}, an Android-based mobile application that
provides location-based service using crowdsourced WiFi-quality
reports. For the place-aware mode, we set the learning duration
$\Delta_L= 1$ hour and the threshold of place detection
$\Delta=0.2$. For the $k$-anonymity mode, we set $k=30$ and
$D_{max}=50$ kilometers which is the longest distance from the
east-most Singapore to the west-most Singapore. When a smartphone
connects to a WiFi access point, it will start to submit
WiFi-quality reports every 40 minutes in the format mentioned in
\Sec{subsec:PrivacyExposureProblem}. The WiFi-Scout crowdsourced
dataset is downloaded from the WiFi-Scout back-end server when the
mobile application is first launched. We conduct two experiments
to study how our algorithm affects privacy exposure and how human
mobility patterns affect privacy exposure. The duration of the
experiment is one week.

\subsection{Experimental Results}
First, we compare our algorithm against the naive sensing
algorithm which submits reports straight without any cloaking when
the smartphone connects to WiFi. In this experiment, a single user
carries two smartphones to collect data, where smartphone 1 is
running the naive sensing algorithm and smartphone 2 is running
our algorithm. \Figss{Fig:ExpResults}(a)-(b) show the results,
where the green markers stand for the real reports and the red
marker stand for the extra redundant reports selected from the
WiFi-Scout crowdsourced dataset using our $k$-anonymity mode. As
it can be seen, the reports submitted by smartphone 1 result in an
extreme high privacy exposure and the user's activity hotspots are
exposed straight since there is no place learning and detection to
avoid exposing long-stay points and the submitted reports are not
cloaked. Note that smartphone 2 submits only 2 real reports since
it learns the user's private points and further avoids submitting
reports when the user stays in these private points. Compared with
the naive sensing algorithm, our algorithm successfully hides the
activity hotspots and improves the privacy exposure by $71.8\%$.

In the second experiment, we consider three users with different
mobility patterns to see how human mobility affects privacy
exposure. We select the three users based on their mobility
diversity in the spatial domain and discuss if our algorithm can
cloak the privacy-sensitive information of users who have seldom
mobility. User 1 and user 2 have less mobility, and their activity
hotspots are located at west and east Singapore, respectively, and
the mobility of user 3 is high as she travels around the whole
Singapore. As shown in \Fig{Fig:ExpResults}(b),
\Fig{Fig:ExpResults}(c), and \Fig{Fig:ExpResults}(d), the reports
submitted by user $1$, user $2$, and user $3$ result in privacy
exposure of $0.255$, $0.241$, and $0.426$, respectively. This
clearly shows that our proposed algorithm effectively reduces
privacy exposure through cloaking users' activity hotspots and
activity transitions, and this can be achieved even at low
user-mobility levels.

\section{Conclusion}\label{Sec:conclusion}
This paper addresses two privacy threats, namely activity hotspot
disclosure and activity transition disclosure, for mobile LBSs. We
have proposed a metric to quantify privacy exposure which
incorporates activity coverage and activity uniformity to evaluate
the privacy exposing levels given a set of user location-tagged
data. We have also defined a privacy exposure problem and proposed
a privacy-preserving algorithm to minimize the exposure of
activity hotspots and activity transitions of users. We have
implemented our algorithm both in a simulation program and on an
Android-based WiFi advisory system, and carried out experiments in
the real world. Both simulation and experimental results
demonstrate that (1) the proposed metric of privacy exposure can
properly characterize different user activity patterns and (2) the
proposed algorithm can effectively reduce privacy exposure by
cloaking users' activity hotspots and activity transitions.

\section{Acknowledgment}
\noindent This work was supported in part by A*STAR Singapore
under SERC grant 1224104046.
\parpic{\includegraphics[width=0.23\linewidth,clip,keepaspectratio]{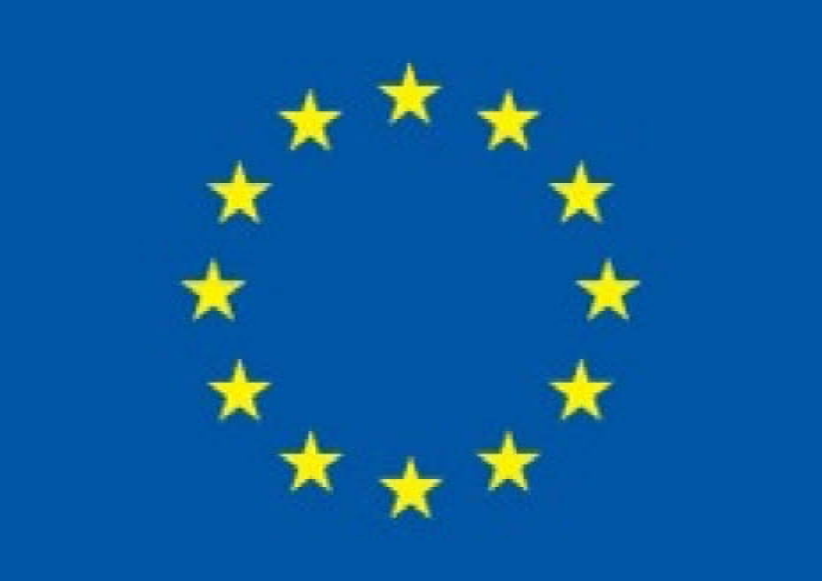}}
\noindent This work has been partially funded by the European
Union's Horizon 2020 research and innovation programme within the
project ``Worldwide Interoperability for SEmantics IoT" under
grant agreement Number 723156.

\bibliographystyle{IEEEtran}                                %


\end{document}